\documentclass[journal=langd5,manuscript=article]{achemso}

\usepackage[version=3]{mhchem} 
\usepackage{subfigure} 
\usepackage{float}
\usepackage[dvipsnames,usenames]{color}
\usepackage{booktabs}
\usepackage{soul}
\usepackage{xcolor}
\usepackage{amssymb}
\usepackage{inputenc}
\usepackage[colorinlistoftodos]{todonotes}
\usepackage[normalem]{ulem}

\usepackage{xr}
\makeatletter
\newcommand*{\addFileDependency}[1]{
  \typeout{(#1)}
  \@addtofilelist{#1}
  \IfFileExists{#1}{}{\typeout{No file #1.}}
}
\makeatother

\newcommand*{\myexternaldocument}[1]{
    \externaldocument{#1}
    \addFileDependency{#1.tex}
    \addFileDependency{#1.aux}
}

\myexternaldocument{supplement}

\title{
Interactions Between a Responsive Microgel Monolayer and a Rigid Colloid: From Soft to Hard Interfaces 
}

\author{Steffen~Bochenek}
\affiliation{Institute of Physical Chemistry, RWTH Aachen University, Landoltweg 2, 52056 Aachen, Germany}
\author{Cathy~E. McNamee}
\affiliation{Department of Chemistry and Materials, Faculty of Textile Science and Technology, Shinshu University, Tokida 3-15-1, Ueda, Nagano, 386-8567, Japan}
\author{Michael Kappl}
\affiliation{Max Planck Institute of Polymer Research, Ackermannweg 10, 55128, Mainz, Germany}
\author{Hans-Juergen Butt}
\affiliation{Max Planck Institute of Polymer Research, Ackermannweg 10, 55128, Mainz, Germany}
\author{Walter Richtering}
\affiliation{Institute of Physical Chemistry, RWTH Aachen University, Landoltweg 2, 52056 Aachen, Germany}
\alsoaffiliation{JARA-SOFT, 52056 Aachen, Germany}

\date{\today}

\begin{document}
\maketitle

\begin{abstract}
\noindent {\textbf{A peer reviewed and extended version of this preprint can be found under S.~Bochenek, C.~E.~McNamee, M.~Kappl, H.-J.~Butt, and W.~Richtering, \textit{PCCP}, 2021, DOI: https://doi.org/10.1039/D1CP01703A.} Responsive poly-\emph{N}-\-iso\-pro\-pyl\-acryl\-amide-based microgels are commonly used as mod\-el colloids with soft repulsive interactions. It has been shown that the microgel-microgel interaction in solution can be easily adjusted by varying the environmental parameters, \emph{e.g.}, temperature, pH, or salt concentration.
Furthermore, microgels readily adsorb to  liquid-gas and liquid-liquid interfaces forming responsive foams and emulsions that can be broken on-demand. 
In this work, we explore the interactions between microgel monolayers at the air-water interface and a hard colloid in the water. Force-distance curves between the monolayer and a silica particle were measured with the Monolayer Particle Interaction Apparatus. The measurements were conducted at different temperatures and lateral compression, \emph{i.e.}, different surface pressures. The force-distance approach curves display long-range repulsive forces below the volume phase transition temperature of the microgels. Temperature and lateral compression reduce the stiffness of the monolayer. The adhesion increases with temperature and decreases with a lateral compression of the monolayer. When compressed laterally, the interactions between the microgels are hardly affected by temperature, as the directly adsorbed microgel fractions are nearly insensitive to temperature. In contrast, our findings show that the temperature-dependent swelling of the microgel fractions in the aqueous phase strongly influences the interaction with the probe. The microgel monolayer changes from a soft to a hard repulsive interface.}

\end{abstract}

\begin{figure}[ht!]
\includegraphics[width=\textwidth]{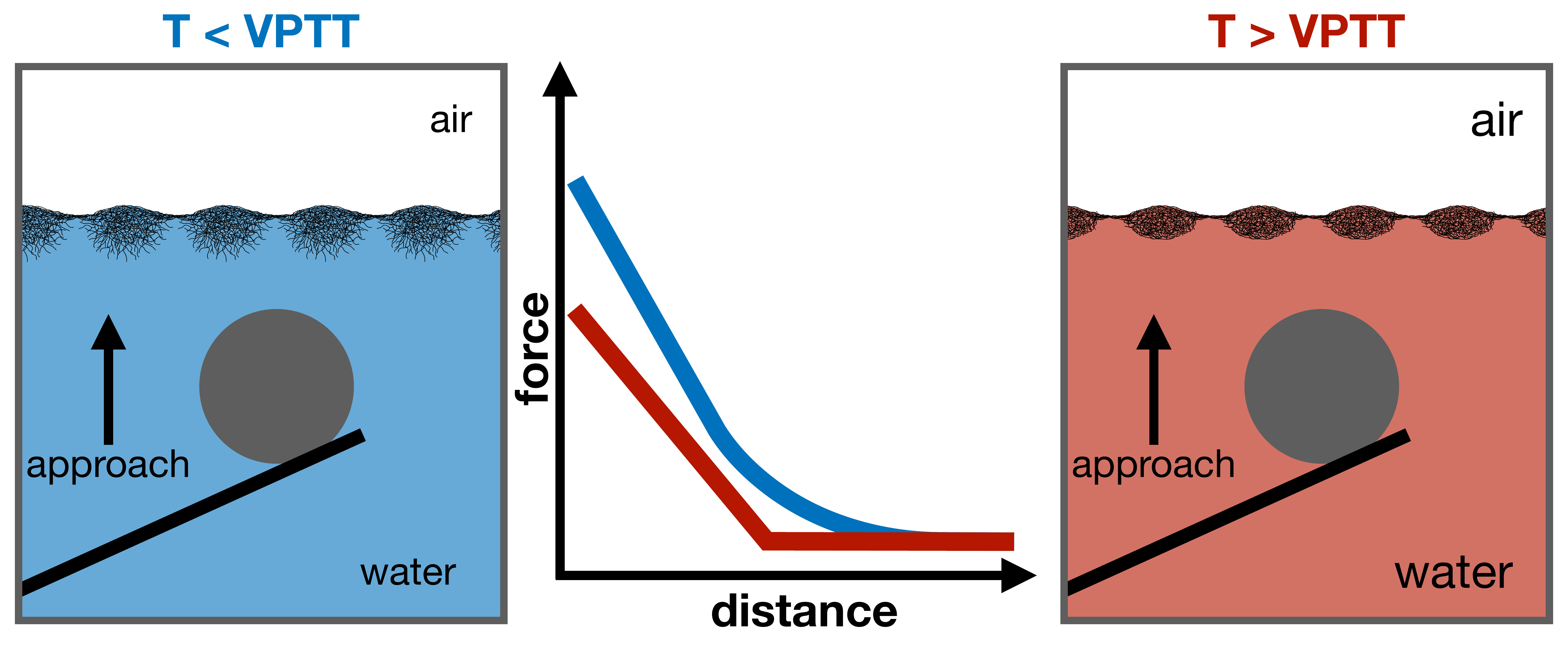}
\caption{TOC}
\label{fig:figure_TOC}
\end{figure}

\section{Introduction.}

Microgels are polymeric particles whose networks are swollen in a good solvent. When dispersed, microgels take up the given solvent to a multitude of their dry volume. Changing the solvent quality, \emph{i.e.}, the microgel-solvent interactions, leads to a (partial) expulsion of the solvent. This influences the properties of the dispersed microgels, such as size, structure, colloidal stability, and interaction potential. The most studied microgels are composed of poly-\emph{N}-\-iso\-pro\-pyl\-acryl\-amide (pNIPAM), which readily disperses in aqueous solution and is thermo-responsive. \cite{Pel00} Typically, their polymer network is cross-linked by \emph{N,N'}-\-methyl\-ene\-bis\-acryl\-am\-ide (BIS) which is consumed faster during the synthesis. \cite{Kro17} For that reason, the microgels have a dense-core-fuzzy-shell structure with a decreasing polymer content and cross-linker concentration from the center to the periphery. \cite{Sti04FF}

Responsive microgels are used for many applications in which the material properties have to be varied \emph{in situ}. For example, the swelling and deswelling of the microgels can be exploited for targeted release of drugs \cite{Swi20, Gue17,Bus17}, responsive micro-reactors \cite{Bor18}, adaptive viscosity modifiers and lubricants \cite{And19}, or sensors \cite{Dan19, Wei19}. In fact, microgels have another remarkable property: a high interfacial activity. pNIPAM-based microgels efficiently lower the surface tension \cite{Zha99,Ngai05} and stabilise emulsions, \cite{Bru09,Ngai05,Ngai06,Fuj05} foams \cite{Dup08,Hor18,Wud18,Fam15}, or coat solids \cite{Ish99,Kes18,MFS19}. Their responsiveness is imparted to the interface, which leads to emulsions and foams that can be broken on-demand. \cite{Ngai05,Fuj05,Dup08,Fam15}

Although the formation and use of responsive emulsions using microgels is reported regularly, the answer to one of the most crucial question remains: Why do microgel-stabilised emulsions become unstable when changing the environmental conditions, such as temperature or pH? Indeed, this question is not trivial. There might be more than one answer, due to the complexity of the emulsion (de-) stabilization \cite{Tad13book} and the many different chemical compositions \cite{Kee07,Ima10}, architectures \cite{Men09, Nic19}, and stimuli-responsivenesses \cite{Chi86,Tan82,Mer15} of microgels. So far, many explanations have been proposed in the literature. One of the first theories includes the desorption of microgels from the interface into the oil-phase due to an increase in their hydrophobicity and a consequent loss of interfacial activity. \cite{Ngai05,Ngai06} However, it has been shown that microgels are surface active at temperatures above the volume phase transition temperature (VPTT) or at different pH values. \cite{Mon10, Gei14} Another explanation attributes the destabilization to the size change of adsorbed microgels within the interface: deswelling in the lateral and vertical directions reduces the microgel-covered interfacial area, which reduces the stability of the emulsion.\cite{Ngai05,Ngai06} Experiments and simulations have provided arguments for and against this theory. \cite{Gei14, Har19, Sch20, Boc19, Boc20, Mae18, Rav17} Moreover, the change of the visco-elastic properties of the microgel monolayers due to the change in the environment has been proposed as a reason for the demulsification. \cite{Bru08,Des14} Indeed, in solution, the swelling state of the microgels has a pivotal influence on their microgel-microgel interaction and the resulting flow properties. \cite{Sen99,Dal08,Sco20} 

In general, the stability of emulsions is determined by the interfacial tension, as well as the mechanical properties and the repulsive electrostatic and/or steric interactions between the films consisting of the emulsifying agents at the interfaces. The aforementioned explanations are focused on the mechanical properties of the monolayers and the decrease in the surface tension.
Recently, Harrer \emph{et al.} discussed that the collapse of the dangling polymer chains of the microgel fraction in the aqueous phase may cause the steric repulsion between the microgel-covered emulsion droplets to be minimized. The swelling state of the microgels normal to the interfaces may also contribute to the destabilization mechanism.

In this work, we aimed to elucidate the effect of the swelling state and compression on the interaction forces between microgel-covered interfaces. We used the simplified model system of a microgel monolayer at an air-water interface and a rigid colloid in the aqueous phase. In order to measure the interactions, we used the Monolayer Particle Interaction Apparatus (MPIA), which is based on the technique of atomic force microscopy. \cite{Gil05} The Langmuir-trough part of the MPIA allows the surface pressure and temperature to be controlled. The MPIA has already been successfully used to measure the interaction between different probes in aqueous solutions and films of surfactants, soft particles, or rigid particles at air-aqueous interfaces as a function of pH, salt concentration and lateral compression. \cite{Mcn10b,Mcn11,Mcn16, Mcn10a, Mcn18} 

We investigated two pNIPAM-based microgels with different volume phase transition temperatures (VPTT). The different VPTTs were achieved by copolymerizing \emph{N,N}-\-di\-ethyl\-acryl\-amide (DEAAM) and NIPAM in one of the microgels. \cite{Kee07} Both microgels were slightly positively charged and had the same size and structure in solution. Compression isotherms and depositions showed that the microgels had the same two-dimensional phase behavior. We exploited the different VPTT to relate the experimental results to the swelling state of the microgels.

Although air-water and oil-water interfaces are different environments, it has been shown that microgels show a similar behavior at the two interfaces. \cite{Boc20} Therefore, we discuss our result in the context of emulsion stabilization. The results give strong evidence that the deswelling of the microgels fractions in the aqueous phase has a major influence on the interaction of microgel-covered interfaces. For the mechanism of demulsification of microgel-stabilized emulsions, not only the strength of the interfacial microgel layer, \cite{ Gei14, Sch20, Boc19, Boc20, Mae18, Rav17, Bru08, Des14} but also the interaction forces between the microgel-covered droplets need to be taken into account. 

\section{Experimental.}

\subsection{Materials.}

\emph{N}-\-iso\-pro\-pyl\-acryl\-amide (Acros Orga\-nics, Belgium),
\emph{N,N}-\-Di\-ethyl\-acryl\-amide (Polysciences Inc., USA),
\emph{N,N'}-\-methyl\-ene\-bis\-acryl\-am\-ide (Alfa Aesar, USA),
\emph{N}-(3-aminopropyl) meth\-acryl\-amide hydro\-chloride (Polysciences Inc., USA),
2,2'-azobis(2-methylpropionamidine) dihydrochloride (V50) (Sigma-Aldrich, USA), chloroform (CHCl$_{3}$, 99.0~$\%$ purity, Wako Pure Chemical Industries, Japan), methanol (MeOH, HPLC grade, Fisher Scientific, Germany), ethanol (EtOH, JIS Special Grade, Wako Pure Chemical Industries, Japan), aqueous ammonia solution (NH3(aq), 25 wt$\%$, analytic grade, WTL Laborbedarf GmbH, Kastellaun, Germany), and N-trimethoxysilylpropyl-N,N,N -trimethylammonium chloride (hydrophilpos, 50~$\%$ in
methanol, ABCR, Germany), and
cetyltrimethylammonium bromide (CTAB) (Fluka Biochemica, Switzerland)
were used as received.
For all the interface experiments ultra pure water (Astacus$^2$, membraPure GmbH, Germany) with a resistivity of 18.2 MOhm$\cdot$cm was used as the sub-phase. To facilitate spreading, propan-2-ol (Merck KGaA, Germany) was used.

\subsection{Synthesis.}

pNIPAM-co-DEAAM and pNIPAM-based microgels were synthesized by precipitation polymerization using the following feed parameters and instructions. For pNIPAM-co-DEAAM microgels the mono\-mers, NIPAM (2.0069~g), DEAAM (3.4854), BIS (0.3748~g), and APMH (0.1384~g) were dissolved in 330~mL double-distilled water. For pNIPAM microgels the mono\-mers, NIPAM (5.4546~g),  BIS (0.3398~g), and APMH (0.1474~g), were dissolved in 330~mL double-distilled water. Both microgels have small amount of the comonomer APMH ($\approx$ 2~mol\%), which allows for post modification of the microgels, such as covalent labeling with fluorescent dyes.\cite{Lyo11, Gel16} The primary amine is positively charged at neutral pH. Therefore, the cationic initiator (V50) and cationic surfactant (CTAB) were used to prevent aggregation. The monomer solution was heated to 65$^\circ$\,C and purged with nitrogen under constant stirring (270~rpm). Simultaneously, the initiator and CTAB (for pNIPAM-co-DEAAM microgels 0.2151~g and 0.0251~g, respectively; for pNIPAM microgels 0.2253~g and 0.0340~g, respectively) were each dissolved in 20~mL water in separated vessels and degassed for one hour. The surfactant-solution was injected into the reaction vessel and stirred for 30 more minutes to equilibrate. The polymerization was initiated by adding the initiator solution in one shot to the reaction flask. The reaction was carried out for 4~h at 65$^\circ$\,C under constant nitrogen flow and stirring. The obtained microgels were purified by threefold ultra-centrifugation at 30~000~rpm (70~000 RCF) and subsequent re-dispersion in fresh, double-distilled water. For the pNIPAM-co-DEAAM microgels, the centrifugation was carried out at T~=~10$^\circ$\,C, in order to assure the microgels were fully swollen. Lyophilization was applied for storage. 

\subsection{Monolayer Particle Interaction Apparatus.}
A micro-manipulator (Model MMO-202D, Narishige) and a  light microscope (BX51, Olympus) were used to attach a silica particle (nominal diameter = 6.8~$\mu$m, Bangs Laboratory, Fishers, USA) to a gold-plated Si$_3$N$_4$ cantilever (V-shaped, nominal spring constant k = 0.15 N/m, OTR8-PS-W, Olympus) using an epoxy resin (Araldite Rapid).  The silica probe was next modified with hydrophilpos using the method reported elsewhere. \cite{Mcn17} 

The forces between the microgel films at the air-water interface and the colloidal probe in water were measured using the MPIA. The MPIA combines a Langmuir trough (Riegler \& Kirstein GmbH, Potsdam, Germany) and a force measurement unit. Detailed information about the MPIA can be found in the literature. \cite{Mcn11} 

The Langmuir trough was cleaned using CHCl$_{3}$ and EtOH. The colloidal probe was then attached to the cantilever holder, water added to the trough, and the water surface suctioned cleaned. The temperature of the water sub-phase was controlled by circulating thermostated water through the base of the trough using a circulation system (C25P, ThermoHaake, Karlsruhe, Germany). Next, a clean mica substrate was placed across the edges of the Langmuir trough filled with water, and force curves were measured between the probe in the aqueous phase and the mica. The calibration factor (CF$_{mica}$), which was needed to convert raw force curves (force [V] versus piezo position [nm] curves) to calibrated force curves (force [nN] versus piezo position [nm] curves), was calculated from the linear contact region of the approach force curves.  After the calibration, the water in the trough was removed and new water added, a minimum of 30 min allowed for water to reach the desired temperature, and then the water surface suctioned cleaned. The microgel monolayer was spread at the air-water interface, 10 min allowed for the spreading solvent to evaporate, and the films compressed to the desired surface pressure. The forces were then measured between that film and the same probe that was used to measure CF$_{mica}$ at an approach/retract velocity of 33 $\mu$m s$^{-1}$ (scan rate: 0.66 Hz, scan size (in z-direction): 25 $\mu$m). At least 50 force curves were measured at each surface pressure.

CF$_{mica}$ was used to convert the raw force curves measured between the particle film and the probe to calibrated force curves. Zero force was defined at large cantilever-film separations, where no surface forces acted on the cantilever. Zero distance was defined as the intersection of the slope of the linear contact region (the area where the probe was in contact with the mica substrate or with the air-aqueous interface in the absence or presence of the microgel monolayer) to the zero force line. 
The adhesion ($F_{ad}$) between the probe and the air-water interface in the absence or presence of a film of microgels at a given surface pressure was taken as the average of the adhesive force measured in all the retract force curves. The effective stiffness ($S_N$) of the air-water interface in the absence or presence of a microgel monolayer was determined from  $S_N$ =  $S_1$/  $S_2$. Here,  $S_1$ is the averaged slope of the linear contact region recorded for all the approach force curves measured between the probe and the air-water interface. $S_2$ is the average of the slope of the linear contact region recorded for all the approach force curves between the probe and the mica substrate, which were used to determine the calibration factor CF$_{mica}$.

\subsection{Compression Isotherms and Depositions.}
Combined compression isotherms and depositions, \emph{i.e.}, gradient Langmuir-Blodgett depositions, were conducted at the air-water interface in a customized liquid-liquid Langmuir-Blodgett trough (KSV NIMA, Biolin Scientific Oy, Finland) made of poly(oxymethylene) glycol. The surface pressure ($\Pi$) was probed with a highly porous platinum Wilhelmy plate (perimeter = 39.24 mm, KSV NIMA, Biolin Scientific Oy, Finland). The plate was attached to an electronic film balance (KSV NIMA, Biolin Scientific Oy, Finland) and placed parallel to the barriers. 

Before each measurement, the trough was cleaned and a fresh air-water interface was created. For temperature control, the trough was connected to an external water bath. Water was added to the trough and the trough was tempered to the desired temperature. A plasma-cleaned rectangular piece of ultra-flat silicon wafer ($\approx$ 1.1 x 6.0 cm, P\{100\}, NanoAndMore GmbH, Germany) was mounted to the substrate holder with an angle of $\approx25^\circ$ with respect to the interface. The substrate was immersed in the water and the water surface suctioned cleaned. After temperature equilibration, the microgel solution (5~mg~mL$^{-1}$ in 50~$\%$ v/v water-propan-2-ol) was added to the interface. The barriers were closed ($v$ =~6.48 cm$^2$ min$^{-1}$) in order to increase the interfacial concentration of the microgels. Simultaneously, the substrate was raised ($v$ = 0.15~$\pm$~0.004~mm min$^{-1}$) through the air-water interface. Compression isotherms and depositions were conducted at $(10.0\pm0.5)~^\circ$C, $(20.0\pm0.5)~^\circ$C and $(40.0\pm0.5)~^\circ$C. A detailed description of gradient Langmuir-Blodgett depositions can be found elsewhere.\cite{Rey16}

\subsection{Atomic Force Microscopy.}

Atomic force microscopy (AFM) measurements of the microgels in the dry state, \emph{i.e.}, at the solid-air interface, were performed using a Dimension Icon with a closed loop (Veeco Instruments Inc., USA, Software: Nanoscope 9.4, Bruker Co., USA). The images were recorded in tapping mode using OTESPA tips with a resonance frequency of 300~kHz, a nominal spring constant of 26 N~m$^{-1}$ of the cantilever and a nominal tip radius of $<$ 7~nm (NanoAndMore GmbH, Germany). 
\subsection{Image Analysis.}

The open-source analysis software \textit{Gwyddion} 2.54 was used to process the AFM images. All images were flattened to remove the tilt and the minimum value was fixed to zero height. The processed AFM micrographs were analyzed with a custom-written MATLAB script\cite{Boc19}. A Delaunay triangulation and Voronoi tesselation were used to compute their nearest neighbor connections. A more detail description can be found in Ref.\cite{Boc19}.

\subsection{Dynamic Light Scattering.}

The hydrodynamic radius of the microgels was measured \emph{via}
Dynamic light scattering (DLS). A laser a with vacuum wavelength of $\lambda_0 = 633$~nm was used. 
The diluted suspensions of the microgels in water have a refractive index of $n({\lambda_0}) = 1.33$. For pNIPAM-co-DEAAM microgels the temperature increment was between T =~(6.0$\pm$0.1)$^\circ$\,C to T = (40.0~$\pm$~0.1)$^\circ$\,C and for pNIPAM microgels between T =~$(15.0\pm0.1)~^\circ$C to T = (49.0~$\pm$~0.1)$^\circ$\,C. The measurements were conducted in 2$^\circ$\,C steps. A thermal bath was filled with toluene to match the refractive index of the glass cuvettes. The scattering angle ($\theta_s$) was varied between 30$^\circ$ and 130$^\circ$, in steps of 20$^\circ$, in order to change the scattering vector $q = 4\pi n/\lambda_0\sin(\theta_s/2)$.
$R_h$ was computed from the diffusion coefficient using the Einstein-Stokes equation. We determined the VPTT from the inflection point of a logistic ``S" shape function fitted to $R_h$ \emph{versus} T.

\subsection{Small-Angle Neutron Scattering.}

Small-angle neutron scattering (SANS) experiments were performed at the KWS-2 instrument (Heinz Maier-Leibnitz Zentrum, Garching, Germany). The scattering vector ($q = 4\pi/ \lambda\sin(\theta_s/2)$, with $\theta_s$ being the scattering angle) was varied by using a wavelength of $\lambda$ = 0.5 and 1~nm for the neutron beam and three sample-detector distances: 20, 8 and 2~m.
The detector was a 2D-$^3$He tubes array with a pixel size of 0.75 cm and a resolution of $\Delta\lambda/\lambda = 10\%$. Data were corrected with sample transmittance and dark count (B$_4$C used). The background, heavy water, was subtracted from all data. The data were acquired at T = 20 and 40$^\circ$\,C and fitted with the Fuzzy-Sphere model.

\subsection{Electrophoretic Mobility and Zeta-Potential.} 

A NanoZS Zetasizer (Malvern Instruments Ltd., England) was used to measure the electrophoretic mobility ($\mu_{el}$) and dynamic light scattering at a scattering angle of 173$^\circ$ (backscatter angle). The temperature range for the pNIPAM-co-DEAAM microgels was between T = 4 to 40$^\circ$\,C, and for the pNIPAM microgels between T = 10 to 50$^\circ$\,C. The measurements were conducted in 2$^\circ$\,C steps for both microgels. A laser with a vacuum wavelength of $\lambda_0$ = 633~nm was used. The Smoluchowski approximation was employed as a model to calculate the zeta potential ($\zeta_{pot}$). The values of $\zeta_{pot}$ are meant to show simply a qualitative trend of the data are shown for comparison with the literature. The temperature of the electrokinetic transition \citep{Lop06} is calculated by fitting a logistic ``S" shape function to the data.

\section{Results and Discussion.} 

The two microgels studied in this work have different VPTTs. The different VPTT are achieved by varying of the chemical composition of the microgels. We used the monomers of NIPAM and DEAAM, which have the same backbone (acryl\-amide), but are differently \emph{N}-substituted. In solution, statistical copolymer microgels of NIPAM ($\approx$ 40~mol\%) and DEAAM ($\approx$ 60~mol\%) have a VPTT of $\approx$ 20$^\circ$\,C, compared to $\approx$ 32$^\circ$\,C of pNIPAM microgels. \cite{Kee07}

\begin{figure}[ht!]
\includegraphics[width=0.5\textwidth]{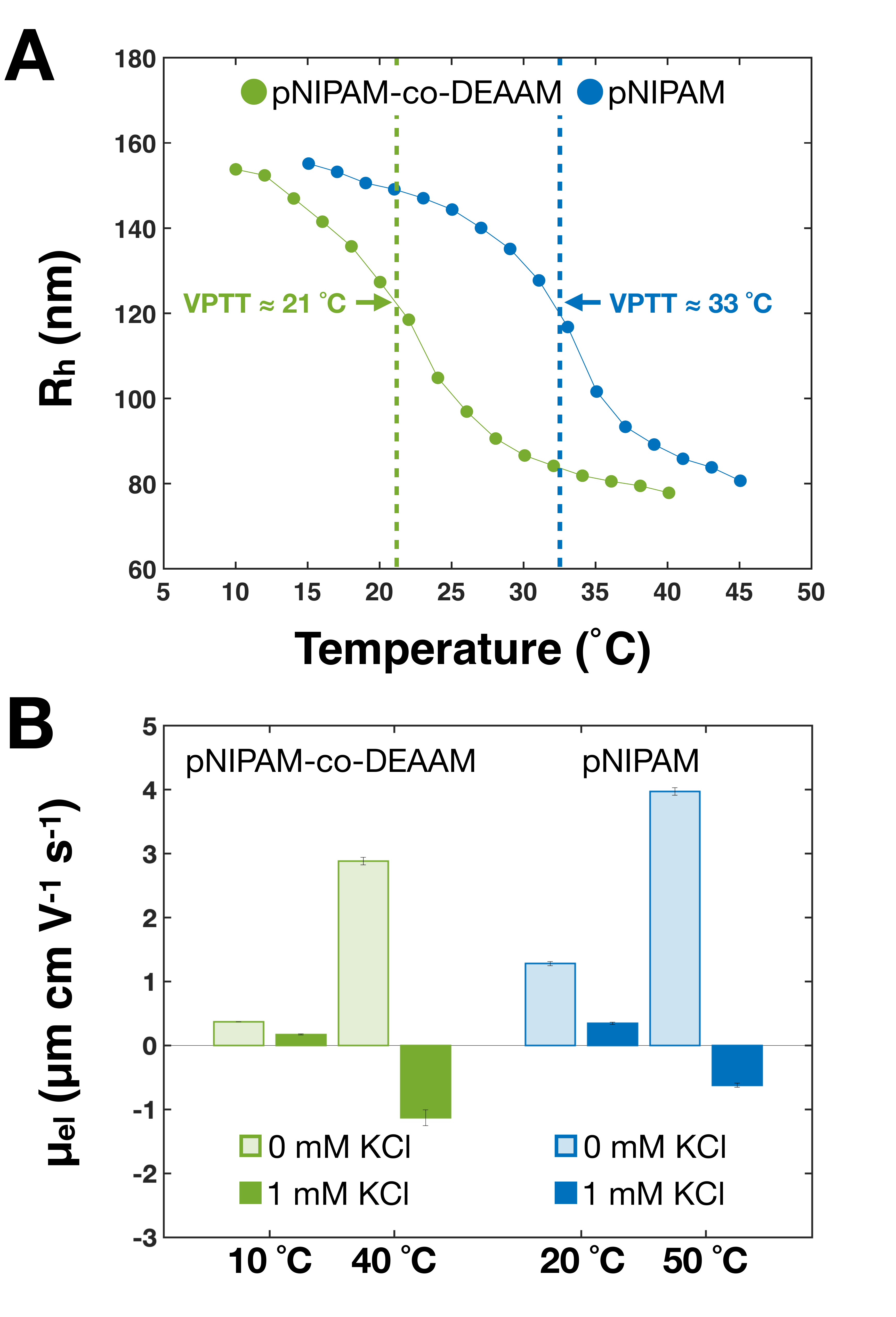}
\caption{Properties of pNIPAM-co-DEAAM and pNIPAM microgels in solutions. (A) Temperature-dependent swelling of the microgels measured with multi-angle dynamic light scattering. (B) Electrophoretic mobility in pure Milli-Q water (0~mM KCl) and with 1~mM KCl as background salt at temperatures below and above the VPTT of the microgels.}
\label{fig:figure_Solution}
\end{figure}

The temperature-dependent swelling of the pNIPAM-co-DEAAM and pNIPAM microgels is shown in Figure~\ref{fig:figure_Solution}A. The results confirm the expected VPTT \cite{Kee07}. In the swollen (T $<$ VPTT) and deswollen state (T $>$ VPTT), the microgels have nearly the same $R_h$ of 160 and 80~nm, respectively. The microgels display the typical inhomogeneous internal structure \cite{Sti04FF} with a decreasing amount of polymer and cross-linker from the center to the periphery (Fig. S1 and Ref.~\cite{Boc19}). Both microgels are monodisperse with size polydispersities calculated from SANS data of (8 $\pm$ 1)~\% for the pNIPAM-co-DEAAM and (7 $\pm$ 1)~\%  for the pNIAPM microgels.

In gerneral, microgels are colloidally stable in solution due to steric and electrostatic repulsions. Below the VPTT, steric repulsions between dangling polymer chains of microgels leads to their colloidal stability. At T $>$ VPTT, electrostatic repulsions between the charged moities 
provides the stability. \cite{Pel00} The electrokinetic characteristics of the pNIPAM-co-DEAAM and pNIPAM microgels were studied with electrophoretic light scattering (ELS). The microgels were synthesized using a cationic initiator (V50) and a small amount of primary amine (APMH). As a consequence, the microgels were positively charged in pure water, as shown by the positive electrophoretic mobility (Fig.~\ref{fig:figure_Solution}B). At elevated temperatures, the microgels displayed a higher $\mu_{el}$ in purified water. 

Analogous to the VPTT, we determined the temperature of the electrokinetic transition from the electrophoretic mobility curves in pure water (Fig.~S2A and B). This transition takes place $\approx$ 2\,K above the VPTT for both microgels. This difference indicates that the strong increase in $\mu_{el}$ with temperature is not only associated with a reduction in the frictional coefficient due to the deswelling of the microgels, but is also due to an increase in the local charge concentration on the microgels' surface caused by the reorganization of the charged moities. \cite{Lop06,Dal00} 

In the case of the pNIPAM and pNIPAM-co-DEAAM microgels, the addition of 1~mM potassium chloride (KCl) leads to a strong decrease in $\mu_{el}$ below the VPTT and a negative mobility above the VPTT (Fig.~\ref{fig:figure_Solution}B). The electrophoretic mobility of the microgels as a function of the temperature in different electrolyte solutions is presented in the ESI (Figure~S2). DLS measurements in 1~mM KCl show an increase of the microgel radius at elevated temperatures (Fig.~S3). 

Although electrostatic repulsion stabilizes the pNIPAM and pNIPAM-co-DEAAM microgels in purified water, the presence of 1~mM salt is sufficient to screen the charges. The microgels become colloidally instable and aggregate above their VPTT. This result highlights that the microgels contain only a very small proportion of charged groups. The characterization of the two microgel systems in bulk shows that the biggest difference between the two microgel systems is their VPTT in aqueous solution. 

\begin{figure}[ht!]
\includegraphics[width=\textwidth]{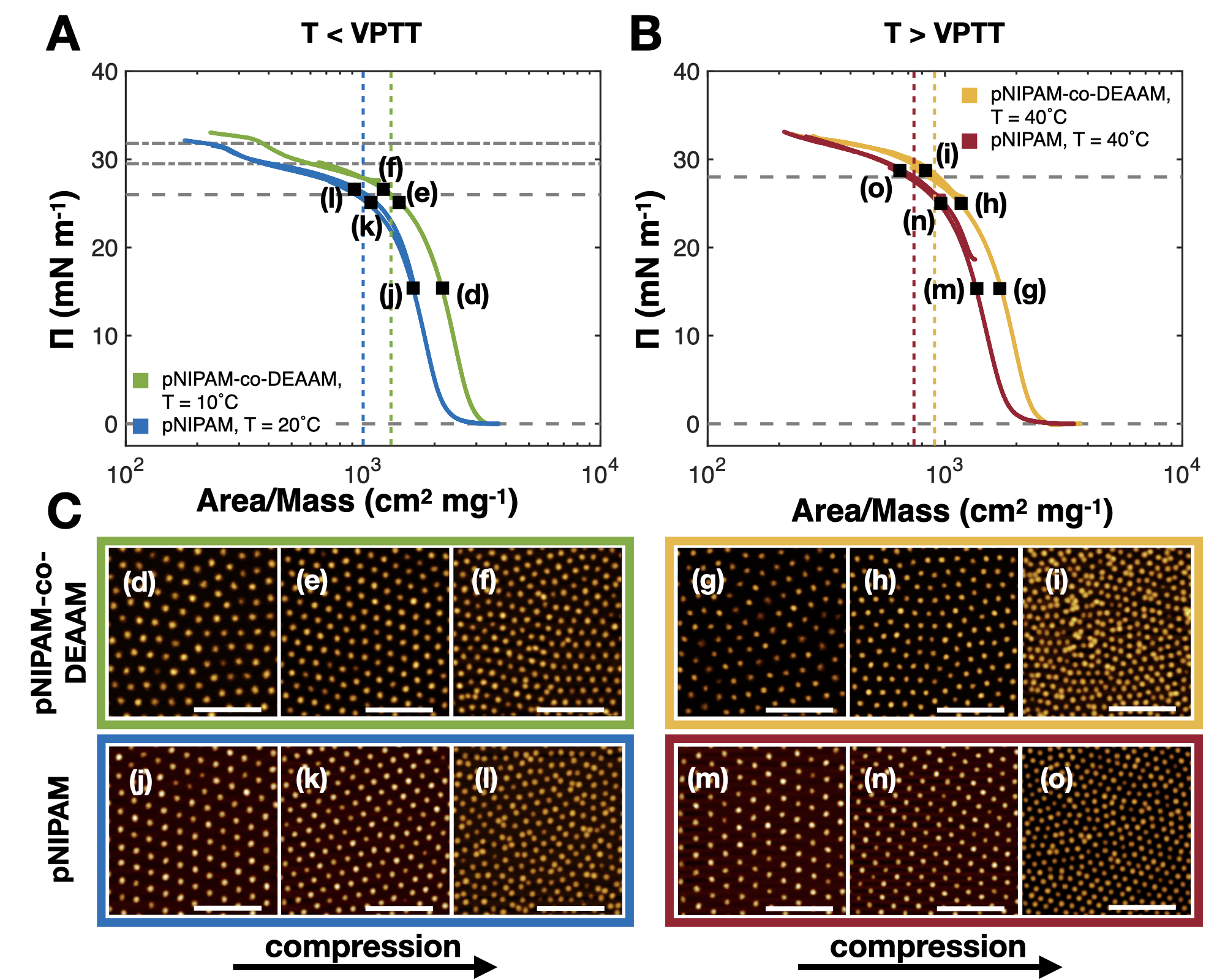}
\caption{$\Pi$ $\emph{versus}$ $Area/Mass$. (A) Monolayers of pNIPAM and pNIPAM-co-DEAAM microgels below their VPTT. (B) Monolayers of pNIPAM and pNIPAM-co-DEAAM microgels above their VPTT. (C) AFM images of the deposited microgel monolayers in dried state. The positions of the images (d-o) on the compression isotherms are highlighted in (A) and (B). Dashed vertical lines represent the onset point of the isostructural phase transition. Gray dashed and dash-dotted horizontal lines highlight the sharp increases in the compression isotherms. Scale bars are 2 $\mu$m.}
\label{fig:figure_structure_MM}
\end{figure}

The compression isotherms at the air-water interface of the monolayers of both microgels are shown in Figure~\ref{fig:figure_structure_MM}A and B. 
$\Pi$ is plotted as a function of the area normalized by the amount of microgel initially added to the interface, \emph{i.e.}, $Area/Mass$. 
We cannot exclude the possibility that a small fraction of the microgels will disperse into the aqueous phase when the microgel solution is spread. Incomplete adsorption may be a problem for microgels with a large amount of charged moities, but is reportedly not a problem for low- or uncharged microgels. \cite{Sch20}
Furthermore, once adsorbed, microgels are considered irreversibly confined at the interfaces due to their high adsorption energy ($\approx$~10\textsuperscript{6}~$k_{\text{B}}T$), which is comparable to solid particles.  \cite{Mont14} Hence, the desorption of microgels into the sub-phase during compression is very unlikely.

The microstructures of the monolayers were investigated using atomic force microscopy images (Figure~\ref{fig:figure_structure_MM}C) after a gradient Langmuir-Blodgett deposition \cite{Rey16} to a silicon wafer. %
Figure~\ref{fig:figure_structure_MM}C shows height images of the dried monolayers at the surface pressures used in the MPIA measurements. The images show that closely packed monolayers were formed in all cases. 
The images were analysed using a custom-written Matlab script to obtain quantitative information about the monolayers  \cite{Boc19}. The values of the number concentrations per area ($N_{area}$) and the center-to-center distances in the first and second crystalline phase ($NND_{1st}$ and $NND_{2nd}$) are summarized in Table~\ref{tab:tab_NND}.

\begin{table}[ht!]
\centering
\resizebox{\textwidth}{!}{\begin{tabular}{cccccc}
    \toprule
     Microgel &  Temperature & $\Pi$ & $N_{area}$ & $NND_{1st}$ & $NND_{2nd}$\\
     - & ($^\circ$\,C) & (mN~m$^{-1}$) & (nm) & (nm) \\
     \midrule
      & 10 & 15 & 4.4 & (507 $\pm$~28) & -   \\
      & & 24 & 5.9 & (438 $\pm$~26) & -  \\
     pNIPAM-co-DEAAM & & 26 & 8.3 & (372 $\pm$~54) & (183 $\pm$~22) \\
      & 40 & 15  & 4.6 & (490 $\pm$~30) & - \\
      &  & 24 & 6.0 & (431 $\pm$~26) & -  \\
      &  & 29 & 14.8 & (296 $\pm$~26) & (191 $\pm$~33) \\
     \midrule
        & 20 & 15 & 4.7 & (489 $\pm$~30) & - \\
        & & 24 & 6.3 & (421 $\pm$~24) & - \\
      pNIPAM & & 26 & 8.1 & (380 $\pm$~35) & (220 $\pm$~20)\\
        & 40 & 15 & 4.8 & (481 $\pm$~21) & - \\
        & & 24 & 6.2 & (431 $\pm$~23) & -  \\
        & & 29 & 15.0 & (275 $\pm$~37) & (195 $\pm$~32) \\

     \bottomrule
     \caption{Number concentration per area ($N_{area}$) and center-to-center distance for the first crystalline phase ($NND_{1st}$) the second crystalline phase ($NND_{2nd}$) of pNIPAM and pNIPAM-co-DEAAM microgels at different compressions and temperatures.}
     \label{tab:tab_NND}
\end{tabular}}
\end{table}

The two-dimensional phase behavior of microgel monolayers has been extensively discussed in the literature, both below \cite{Gei14,Rey16} and above the VPTT. \cite{Boc19, Boc20} Briefly, the softness of the microgels allows their deformation at the interface. Upon adsorption, the contact area of the microgel with the interface is maximised in order to reduce as many unfavourable air- (or oil-) water contacts as possible and to lower the surface tension. The deformation is limited by the cross-links within the microgels' polymer network, which decrease in amount from their center to the periphery \cite{Sti04FF, Kro17}. The resulting shape of the adsorbed microgels is described as a core-corona or ``fried-egg''-like structure. \cite{Bru09,Des11,Gei14,Rey16,Boc19,Fab19} The corona represents the portion of the microgel at and near the interface and the core represents the portion that is still in the aqueous phase.

At sufficiently large concentrations, the microgel coronae contact one another and a densely packed monolayer is formed. The compression isotherms below and above the VPTT show a sharp first increase in the surface pressure (Figures~\ref{fig:figure_structure_MM}A and B, between gray dashed horizontal lines) starting at $\approx$~3000 and 2500~cm$^2$ mg$^{-1}$ for the pNIPAM-co-DEAAM and pNIPAM microgels, respectively. The microgels in the corona-corona contact region have a hexagonal order as shown in Figure~\ref{fig:figure_structure_MM} C, (d), (g), (j), and (m). Lateral compression of the monolayers decreases the lattice constant, \emph{i.e.}, the center-to-center distance, causing the surface pressure to increase. \cite{Boc19} 
At higher lateral compressions, an isostructural phase transition takes place, where two the crystalline phases coexist (AFM images (f), (i), (l), and (o) in Fig.~\ref{fig:figure_structure_MM}). \cite{Rey16, Boc20} The onset of the transition is highlighted by the dashed vertical lines in Figures~\ref{fig:figure_structure_MM}A and B. As the microgels progress into the second crystalline phase, the compression isotherms show a pseudo-plateau where the $\Pi$ becomes nearly independent of $Area/Mass$. The influence of temperature on the microgel monolayers becomes visible at even higher lateral compressions, when all the microgels are in the second crystalline phase. \cite{Boc19, Boc20} Below the VPTT, the second crystalline phase is compressible and the surface pressure increases once more (Fig.~\ref{fig:figure_structure_MM}A, between the gray dash-dotted horizontal lines) before the monolayer collapses. In contrast, at T $>$ VPTT, the second crystalline phase is incompressible and the monolayer collapses without a further strong increase in $\Pi$ (Fig.~\ref{fig:figure_structure_MM}B). The compression isotherms of the microgel monolayers change from two sharp increases in the surface pressure at T $<$ VPTT (Fig.~\ref{fig:figure_structure_MM}A dashed and dash-dotted horizontal lines) to a single increase above the VPTT (Fig.~\ref{fig:figure_structure_MM}B dashed horizontal lines).

\begin{figure}[ht!]
\includegraphics[width=0.8\textwidth ]{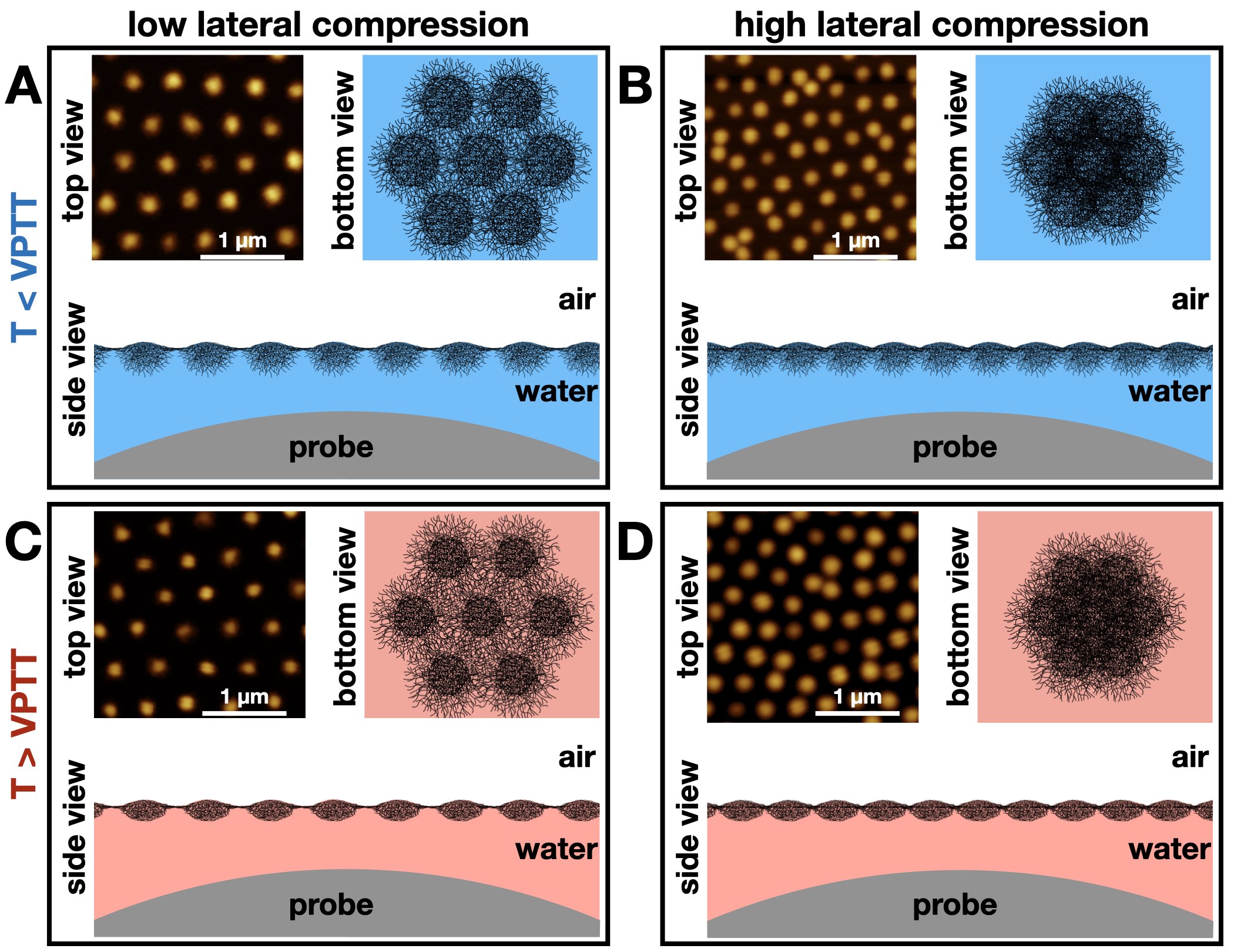}

\caption{Sketches of microgel monolayers at air-water interfaces and AFM images. AFM images show microgel monolayers in dried state (top view). The small drawings show microgel monolayers from the aqueous phase looking at the air-water interface (bottom view). Large sketches display the monolayer from the side (side view). To clarify the size relationship, the probe is sketched at the bottom edge. Microgel monolayers are shown below (A and B) and above (C and D) the VPTT and at low (A and C) and high (B and D) lateral compression. 
}
\label{fig:figure_sketch}
\end{figure}

For the interpretation and discussion of the MPIA experiments below, we also briefly summarize the results for swelling/deswelling of microgels at liquid interfaces. \cite{Mae18, Har19, Boc19, Boc20} For illustration, the microgel monolayers at different temperatures and compressions are sketched in Figure~\ref{fig:figure_sketch} together with the colloidal probe. The crucial difference between the microgels in solution and those at the air-water interface is their altered volume-phase transition when the microgels are adsorbed. \cite{Har19, Boc19, Boc20} The size of the directly adsorbed fractions of the microgels, \emph{i.e.}, the corona, depends on the surface tension and is nearly temperature-independent. In contrast, the cores of the adsorbed microgels have similar properties as microgels in the aqueous bulk phase, \emph{e.g.}, they can still deswell as a function of temperature. \cite{Boc19,Har19,Boc20} The microgel cores are therefore strongly swollen by water and have dangling polymer chains below the VPTT (Figures~\ref{fig:figure_sketch}A and B, side view). At T $>$ VPTT, the cores are deswollen and their dangling polymer chains are collapsed (Figures~\ref{fig:figure_sketch}C and D, side view). \cite{Har19, Boc19} Compared to the swollen state, the deswollen cores not only have a a reduced thickness, but also a smaller (lateral) diameter and a larger polymer volume fraction compared to the swollen state. \cite{Boc19, Mae18} The bottom view in Figure~\ref{fig:figure_sketch}C also shows that by deswelling the cores, more of the firmly adsorbed polymer layer, \emph{i.e.}, the corona, is effectively exposed and accessible from the aqueous sub-phase at the same center-to-center distance. At the onset of the isostructural phase transition, the microgel cores contact one another (Figures~\ref{fig:figure_sketch}B and D, bottom view). \cite{Boc19} As a result, the monolayers are less undulated below and above the VPTT (Figures~\ref{fig:figure_sketch}B and D, side view).

We used the MPIA \cite{Gil05} to measure force-distance curves between microgel monolayers at flat air-water interfaces 
and a colloidal probe. The colloidal probe was funtionalized with hydrophilpos, in order to obtain a hydrophilic, positively charged surface of the probe. \cite{Mcn09} This reduces the attractive electrostatic interactions that existed between the originally negatively charged silica particle and the cationic microgel monolayer and prevented the deposition of the microgels to the probe. Force distance curves were measured between the probe in the Milli-Q water sub-phase and the monolayer at the water surface.  The raw data were converted into force (nN) \emph{versus} distance (nm) and shifted to zero distance (for details see the experimental part). The interactions were measured as a function of temperature (10, 20, 33 and 40$^\circ$\,C) and monolayer surface pressure (15, 24, and 26 or 29~mN~m$^{-1}$).

\begin{figure}[ht!]
\includegraphics[width=0.7\textwidth]{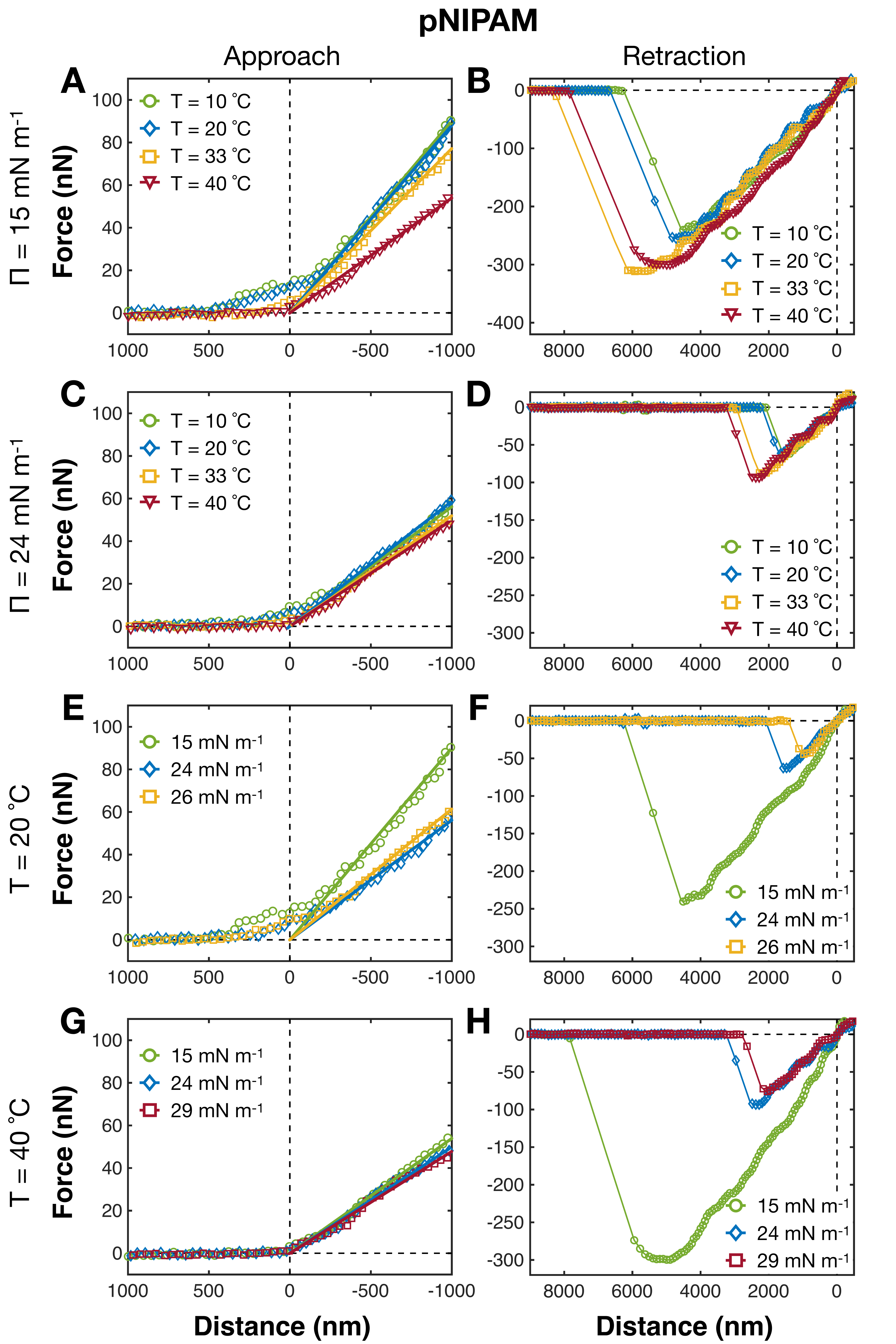}
\caption{Examples of force-distance curves of pNIPAM microgel monolayers at the air-water interface measured with the probe (diameter of 6.84 $\mu$m) from the aqueous phase. Approach (left) and retraction (right) force-distance curves. (A,B) Force distance curves as a function of temperature. (C,D) Force distance curves as a function of compression for T = 20~$^\circ$\,C. (E,F) force distance curves as a function of compression for T = 40 $^\circ$\,C.}
\label{fig:figure_FD_SB104c}
\end{figure}

Examples of the force-distance curves are shown in Figure~\ref{fig:figure_FD_SB104c} for the pNIPAM microgels and in Figure~S4 for pNIPAM-co-DEAAM microgels. The approach curves of both microgels show long-range repulsive interactions at low temperatures (T $<$ VPTT), expressed by a gradual increase in the force. Temperatures close to the microgels' VPTT caused the distance of the onset of the long-range repulsion to significantly decrease; it fully vanished above the VPTT. Increasing the lateral compression had no effect on the long-range repulsion (Figures~\ref{fig:figure_FD_SB104c}E,G and S4E,G). In comparison, the force distance approach curves in Figure~S5 showed an attraction between the probe and the bare air-water interfaces. 
The air-aqueous interface is negatively charged. \cite{tak05,Cha09,Pol20} Therefore, the attraction is explained by electrostatic interactions between the negatively charged bare air-water surface and positively charged probe.

We explain the long-range repulsion between the monolayers and the probe by steric repulsions resulting from the swelling state of microgels' fractions in the aqueous phase. As illustrated in Figure~\ref{fig:figure_sketch}, below the VPTT, the cores of the microgels are hydrated and posses dangling chains that extend into the aqueous phase. The dangling polymer chains and the hydrated cores are compressed normal to the interface during the approach of the probe. 
At elevated temperatures, the cores and dangling chains are collapsed (Fig.~\ref{fig:figure_sketch}C and D). In this case, the probe cannot compress the deswollen cores and the microgel film is deformed or bent. The incompressibility of the deswollen microgel cores was also observed upon lateral compression, \emph{i.e.}, in the compression isotherms (Fig.~\ref{fig:figure_structure_MM}A and B).

The approach curves (Figures~\ref{fig:figure_FD_SB104c} and~S4) showed a linear contact region at negative distances.  
In this constant compliance region, the probe and monolayer cannot come closer together, \emph{i.e.}, they have a constant separation. It is expected that the movement of the colloidal probe deforms the fluid interface, which is the microgel film in this case. \cite{Pit02,Dav14,Ana16,But17} Thus, the slope of the region of constant compliance is proportional to the stiffness of the interface and the spring constant of cantilever. \cite{Duc94, Har99} The slope of the linear contact region decreases across the VPTT of the microgels and with lateral compression of the pNIPAM microgel monolayer.

Examples of the retract curves are shown in Figures~\ref{fig:figure_FD_SB104c}B, D, F, H and S4B, D, F, H. The influence of temperature on the retract force-distance curves is predominately observed across the VPTT of the microgels in solution. Increasing the temperature causes the colloidal probe to adhere stronger to the monolayer. Larger distance and forces are required to separate the probe form the microgel film (Fig.~\ref{fig:figure_FD_SB104c}B, D). Lateral compression of the monolayers yields the opposing result. At higher concentrations of microgels at the interface, \emph{i.e.}, larger $\Pi$-values, the probe can be separated from the monolayer by a weaker force.  The strongest reduction was observed between 15~mN~m$^{-1}$ and 24~mN~m$^{-1}$ for all the temperatures (Fig~\ref{fig:figure_FD_SB104c}F, H). 

\begin{figure}[ht!]
\includegraphics[width=0.7\textwidth ]{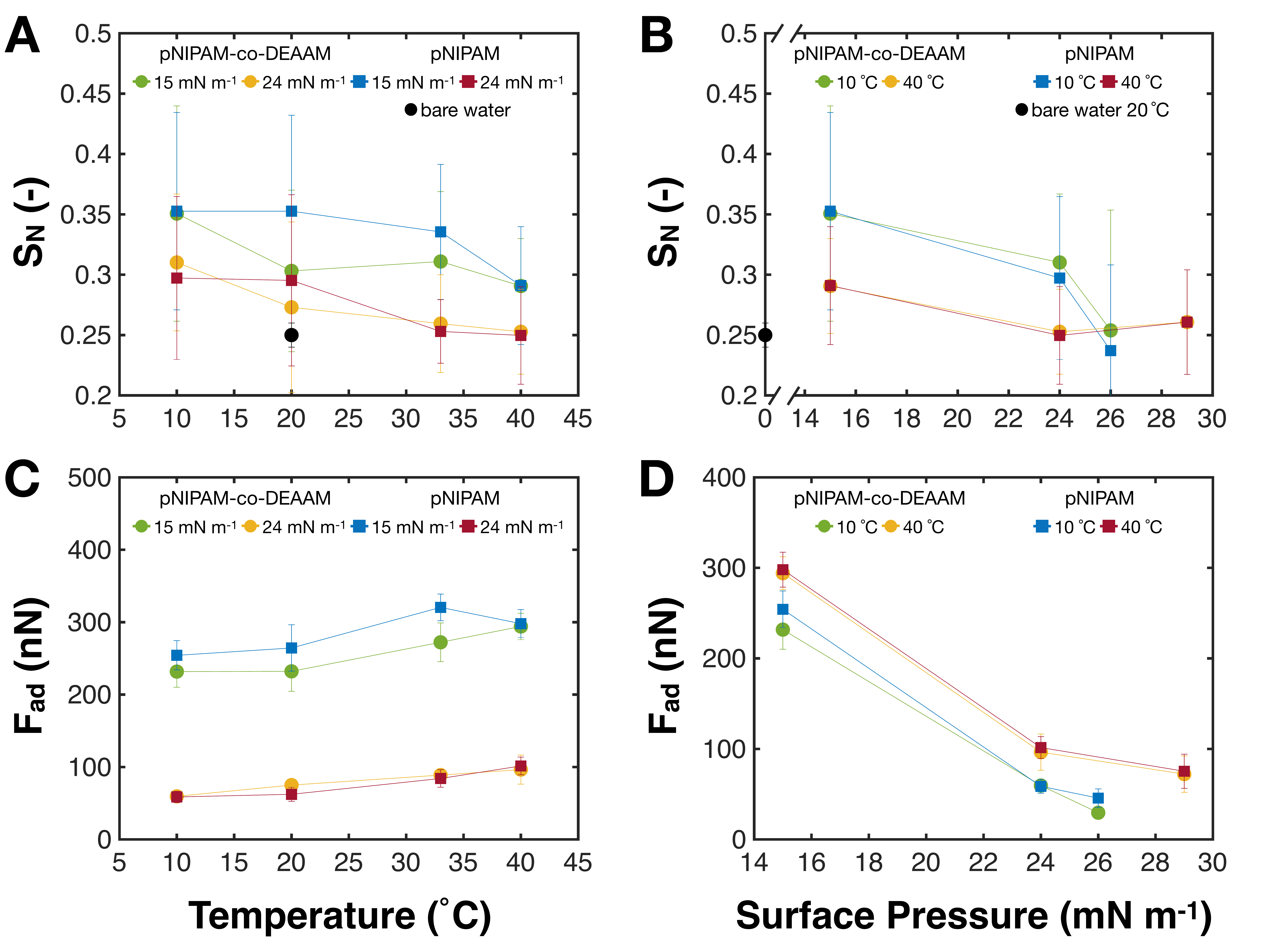}

\caption{Effective stiffness ($S_N$) and adhesion force ($F_{ad}$) of the pNIPAM-co-DEAAM and pNIPAM microgel monolayers. (A) $S_N$ as a function of temperature. The black circle displays the bare water interface. (B) $S_N$ as a function of lateral compression. The black circle displays the bare water interface. (C) $F_{ad}$ as a function of temperature. (D) $F_{ad}$ as a function of lateral compression. In C and D, $F_{ad}$ of the bare water interfaces is out of scale with $F_{ad}$ = (620 $\pm$ 25)~nN. All values are averaged over at least 50 force curves, the standard deviation is given as error.
}
\label{fig:figure_SN_Fad}
\end{figure}

We determined the effective stiffness ($S_N$) and the force of adhesion ($F_{ad}$) of the monolayers from the force distance curves. Figure~\ref{fig:figure_SN_Fad} shows $S_N$ and $F_{ad}$ for pNIPAM-co-DEAAM and pNIPAM microgel monolayers as a function of the temperature (Fig.~\ref{fig:figure_SN_Fad}A and C) and $\Pi$ (Fig.~\ref{fig:figure_SN_Fad}B and D).
$S_N$ was calculated from the region of constant compliance of the approach force distance curves. $S_N$ is the ratio of the slope of the region of constant compliance between the probe and the microgel monolayers and between the probe and a mica substrate (for details see the experimental part). The constant compliance region gives a linear force-distance region. This region was fit by a linear line to give $S_N$. Examples of the fits are given in Figures~\ref{fig:figure_FD_SB104c}A,C,E,G and ~S4A,C,E,G. $F_{ad}$ between the probe and the monolayer was quantified as the difference between the minimum and zero force in the retract force distance curves. Both parameters were averaged over at least 50 force curves. The error bars are a consequence of the variation in the force curves for the same temperature and lateral compression.

$S_N$ describes the deformability of the interface. \cite{Cha01} For bare air-water interfaces, the surface tension is thought to affect the $S_N$, as the effective stiffness of an air-aqueous interface has been shown to decrease with a decreasing surface tension. \cite{Cha01} Figure~\ref{fig:figure_SN_Fad}A shows that the microgel monolayers were stiffer than the bare air-water interface at T = 20~$^\circ$\,C, even though the surface tension was greatly reduced. Thus, additional factors need to be considered for the effective stiffness when a film of particles or surfactants is present. 

Increasing the lateral attraction between particles within the film, \emph{e.g.}, by enhancing capillary forces \cite{Mcn16} or hydrophobic interactions \cite{Mcn14, Mcn11}, leads to stiffer monolayers. In general, $S_N$ of the film is higher for films of particles with a lower lateral mobility. This is because the particles in such a film are moved less by the incoming probe, causing the film to remain more intact rather than a film whose particles have a high lateral mobility. Films with a low lateral mobility may result from films with a high viscosity or films whose particles show strong inter-particle attractions. The higher stiffness of the microgel monolayers compared to the clean air-water interface is in line with expectations from interfacial rheological measurements. Both dilatational and shear rheology have shown that microgel monolayers strongly affect the visco-elastic properties of fluid interfaces. \cite{Zif14,Rey16,Hua17,Mae18, Bru10} This increased viscosity of the microgel-laden interface also impacts the deformability of the monolayer normal to the interface. 

Heating the microgel monolayer tended to decrease $S_N$ (Fig.~\ref{fig:figure_SN_Fad}A). The center-to-center distances of the microgels are virtually the same below and above the VPTT, but the microgel cores deswell in vertical and lateral directions. \cite{Boc19,Mae18,Har19} The decreased thickness of the monolayer, the collapse of the cores, and the dangling polymer chains in the aqueous side lead to larger areas of the interfaces being covered by the thin coronae of the microgels (Fig.~\ref{fig:figure_sketch}A and C, bottom view). This may allow the microgels film to deform or bend easier. 
Similarly, the stiffness of surfactant monolayers has been found to decrease with the chain length of surfactants, \emph{i.e.}, a decrease of the film thickness. \cite{Mcn10a} Moreover, Particle Tracking Microrheology has shown that increasing the temperature leads to a transition from visco-elastic to a viscous fluid, \cite{Mae18} displaying the enhanced mobility of the adsorbed microgels. 

Below and above the VPTT (Fig.~\ref{fig:figure_SN_Fad}B), lateral compression also tends to decrease $S_N$. If we consider our above explanation for the effect of temperature on the monolayer stiffness, the decrease of $S_N$ with $\Pi$ is surprising. The image analysis of the deposited microgels (Table~\ref{tab:tab_NND}) show that the microgels are pushed laterally into each other. This should jam the microgels as illustrated in Figure~\ref{fig:figure_sketch} and drastically decrease their mobility, which should increase $S_N$. In contrast, shear rheology measurements \cite{Rey16} and calculations of the surface elasticity of microgel monolayers \cite{Pin14} show a maximum around 15-20~mN~m$^{-1}$. A similar trend has been observed for other adsorbed species, such as TiO$_2$ particles \cite{Mcn16}, polystyrene particles \cite{Mcn14, Mcn17}, and surfactants \cite{Mcn12,Mcn10a}. The decrease of $S_N$ is explained by the decreased surface tension, that is an increase in surface pressure. At small surface pressures, \emph{i.e.}, at larger center-to-center distance, more force is needed for the deformation. This is because the Laplace pressure of the system is higher, counter acting the increase of the surface area. \cite{Cha01} Consequently, the stiffness is reduced for higher $\Pi$.
 
The presence of microgels at the interface significantly reduces $F_{ad}$ between the probe and the interface. For the bare water interface, $F_{ad}$ is (620 $\pm$ 25) nN. A microgel monolayer at 15~mN~m$^{-1}$ and T = 10~$^\circ$\,C has an $F_{ad}$ of roughly 250~nN (Fig.~\ref{fig:figure_SN_Fad}C). This strong adhesion observed for bare water can be explained by a three-phase contact formation between the probe and the interface. A strong capillary force needs to be overcome, when the probe is removed from the air-water interface. Scheludko $\&$ Nikolov \cite{Sch75} have calculated the force required to pull a sphere out of a liquid ($F_{adh}$). When a particle is moved from air into a liquid, 

\begin{equation}
    F_{adh}~=~2 \pi R \gamma sin^2 \left(\frac{\theta}{2}\right) \notag \,,
    \label{eq:Scheludko}
\end{equation} 

where $R$, $\gamma$, and $\theta$ are the radius of the sphere, the interfacial tension of the air-liquid interface, and the advancing contact angle of the sphere with the air-water interface, respectively. Inserting $R$ =3.4~$\mu$m, $F_{ad}$ = 620 nN for $F_{adh}$, and $\gamma$~= 0.072~N~m$^{-1}$ leads to $\theta$~=~79$^\circ$. The reduction in $F_{ad}$ by the presence of the microgels at the air-water interface is explained by i) the microgels prevent the probe from forming a three-phase contact line and ii) there is an electrostatic repulsion between the microgel and the probe, as both are positively charged.



The adhesion between the microgel monolayers and the probe decreases with increasing temperature (Figure~\ref{fig:figure_SN_Fad}C). 
For particle monolayers it was shown that the $F_{ad}$ decreases as a result of: (i) electrostatic repulsion between the monolayer and the probe (if the charges of both have the same sign); and (ii) inhibition of direct contact between the probe and the negatively charged sites of the air-water interface, \emph{i.e.}, the monolayer blocks the attractive electrostatic interactions. \cite{Mcn17} Collapsed thermo-responsive microgels are known to be colloidally stabilized in solution by charged moieties originating, for example, from the ionic initiator fragments. \cite{Pel00} Both pNIPAM and pNIPAM-co-DEAAM microgels showed the expected increase in electrophoretic mobility with temperature in Milli-Q water (Fig~\ref{fig:figure_Solution}B). As discussed above, the cores of adsorbed microgels and the microgels in solution have similar properties. Following this assumption, the aqueous side of the microgels monolayers is expected to become more positively charged with increasing temperatures. Since the probe is also positively charged, $F_{ad}$ should decrease with temperature due to electrostatic repulsion. However, in Figure~\ref{fig:figure_SN_Fad}C the opposite was found. Electrostatic repulsion between the like-charged microgels and the probe therefore seems be of minor importance. Moreover, (dominant) repulsive electrostatic forces caused a long-range repulsion in the approach curves, \cite{Mcn12} which was not observed in Figures~\ref{fig:figure_FD_SB104c}A,C and S4A,C. Therefore, the reduced probe-microgel monolayer adhesion with increasing temperature is explained by the blockage of the negatively charged sites of the air-water interface by the presence of the pNIPAM or pNIPAM-co-DEAAM microgels. The deswelling of the microgel cores leads to larger areas of the interfaces being covered by the thin coronae of the microgels (Fig.~\ref{fig:figure_sketch}A and C, bottom view). This results in a poorer blockage of the negatively charged sites at higher temperatures and an increasing adhesion.



Additionally, hydrophobic interactions between the probe and the monolayer contribute to the adhesion. The contact angle of the hydrophilpos probe calculated from above (79$^\circ$) shows that the hydrophilpos probe is not perfectly hydrophilic. When the probe is retracted from the microgel monolayer, parts of the polymer network that were adsorbed to the probe need to be rehydrated. At elevated temperatures, rehydration is more unfavorable than at low temperatures, due to the increased hydrophobicity of the pNIPAM and pNIAPM-co-DEAAM polymer. In the literature, increasing $F_{ad}$ across the VPTT has been reported for adhesion measurements of pNIPAM coated solid substrates (both brush and microgel systems) in aqueous environment. \cite{Sch10,Kes10}

The decrease in adhesion with increasing $\Pi$ that is shown in Figure~\ref{fig:figure_SN_Fad}D can be attributed to the lateral compression of the microgels. During lateral compression, the center-to-center distance of the microgels decreases (Table~\ref{tab:tab_NND}) and the monolayer becomes denser (Fig.~\ref{fig:figure_sketch}). This causes the attractive electrostatic interaction between the probe and the air-water interface to be blocked more.

The MIPA experiments show the same trends as above, \emph{i.e.}, almost the same $S_N$ and $F_{ad}$ values for the pNIPAM and pNIPAM-co-DEAAM microgel monolayers were obtained. A significant difference was found only in the temperature response of the microgel monolayers. For example, the long-range repulsion disappeared between 20 and 40~$^\circ$\,C for the pNIAPM microgels and between 10 and 33~$^\circ$\,C for the pNIPAM-co-DEAAM microgels, \emph{i.e.}, when the respective VPTT of the microgels is exceeded. This difference illustrates that the changes are related to the deswelling of the microgel fractions on the aqueous side of the interface.

Lastly, we want to discuss the meaning of the MPIA results for the (de-) stabilization of emulsions formed by microgels. As presented in the introduction, one of the most prominent properties of microgels at interfaces is the formation of stimuli-responsive emulsions. A large number of studies on the break-down of microgel-stabilized emulsions have been published in the last decades, providing various explanations. \cite{Ngai05,Ngai06,Bru08,Bru08b,Des11,Mas14,Kwo18b,Kwo19, Pin14,Bru10,Mae18,Mon10} A recent review by Fernandez-Rodriguez \emph{et al.}, \cite{MAF20} summarizes the complexity of microgel monolayers and the abundance of the different aspects.

Here, we probed the interactions by using a model system of a hard colloid in the aqueous phase and a flat microgel monolayer at the air-water interface. Indeed, as suggested by Harrer \emph{et al.} \cite{Har19}, the deswelling of the microgel cores minimizes the steric repulsions between microgel-covered interfaces. The force-distance curves (Figures~\ref{fig:figure_FD_SB104c}A,C and S4A,C) demonstrate that increasing the temperature above the VPTT leads to a loss of long-range repulsion, independent of the lateral compression. The long-range repulsion is correlated to the steric froces between the probe and the monolayer due to dangling chains, which collapse at temperatures above the VPTT. \cite{Har19,Boc19}  Nevertheless, we want to point out that we do not expect the loss of repulsive interaction to be the sole explanation for the emulsion destabilization. We suggest that it is a combination two processes: a loss of steric stabilization and a decrease in the visco-elastic properties of the interface, \emph{i.e.}, strength of the microgel film.

\section{Conclusion.}
We studied the interactions of a probe in the aqueous phase with pNIPAM and pNIPAN-co-DEAAM microgels monolayers using the MPIA. Our results highlight a transition from soft to hard interfaces. Below the VPTT, a long-range soft repulsion normal to the interface exists between the microgel monolayers and the probe due to steric interactions. In the deswollen state, this interaction becomes short-ranged or hard-core like and the adhesion between the monolayer and the probe comes larger. This transition is in contrast to the lateral compression of microgel monolayers reported in the literature. \cite{Boc20} The lateral interaction of microgel monolayers is dominated by the directly adsorbed fractions of the microgels. These fractions are hardly affected by temperature and only at very high interfacial concentrations are the interactions affected by the swelling state of the microgel cores.

The results are discussed in the context of microgel-stabilized emulsions, suggesting that the changes in the interaction normal to the interface may contribute to their break-down. We propose that the mechanism of de-emulsification is a combination of both the reduction of repulsive interactions and a decrease of the visco-elastic properties in the interface. Indeed, it should be considered that a simplified model system of a microgel monolayer at a flat interface and a hard colloidal probe is used. While microgels have nearly identical properties at flat air-water or oil-water interfaces, \cite{Boc20} future MPIA measurements utilizing probes that reflect a more native environment should be conducted. For example, this has been done for rice starch or TiO$_2$ particles. \cite{Mcn16,Mcn18} In solution, intra-microgel interactions, which dominate the deswelling, also lead to inter-microgel aggregation. \cite{Sti04FF,Cra06} Measurements where both the probe and the monolayer are made from the same material, \emph{e.g.} by coating a silica particle with microgels, may further contribute to the understanding of the de-emulsification of microgel-stabilized ``smart'' emulsions.

\section{Acknowledgements.}
C.~E.~M. and S.~B. contributed equally to this work. The authors acknowledge financial support from the SFB 985 ``Functional Microgels and Microgel Systems'' of Deutsche Forschungsgemeinschaft within project B8. This work is based upon SANS experiments performed at the KWS-2 instrument operated by J\"ulich Centre for Neutron Science (JCNS) at the Heinz Maier-Leibnitz Zentrum (MLZ), Garching, Germany.

\section{Conflicts of Interest}
There are no conflicts to declare.

\section{Data Availability}
Additional research data for this article may be accessed at no charge at https://hdl.handle.net
/21.11102/ac75b142-e857-11ea-afb2-e41f1366df48

\bibliography{main_bib}

\end{document}


\newpage
\section{Characterization in Bulk}

\begin{figure}[ht!]
\includegraphics[width=.90\textwidth ]{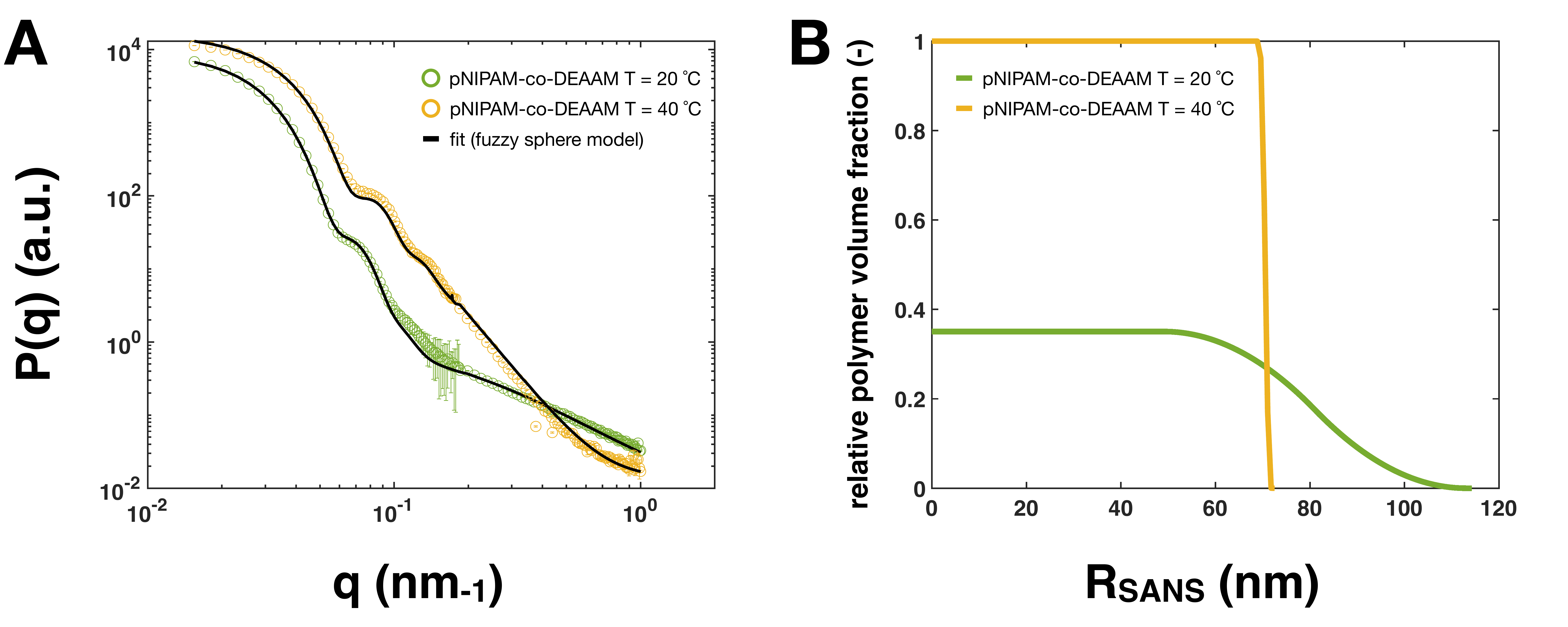}
\caption{Small-angle neutron scattering data and fits of pNIPAM-co-DEAAM microgels. (A) Particle form factor, $P(q)$, $\emph{versus}$ scattering vector, $q$, with fits at T = 20$^\circ$\,C and T = 40$^\circ$\,C. (B) Relative polymer Volume fraction \emph{versus} radius, R, from the fits of the fuzzy-sphere model in (A).  
}
\label{fig:figure_SANS}
\end{figure}

\begin{figure}[ht!]
\includegraphics[width=.90\textwidth ]{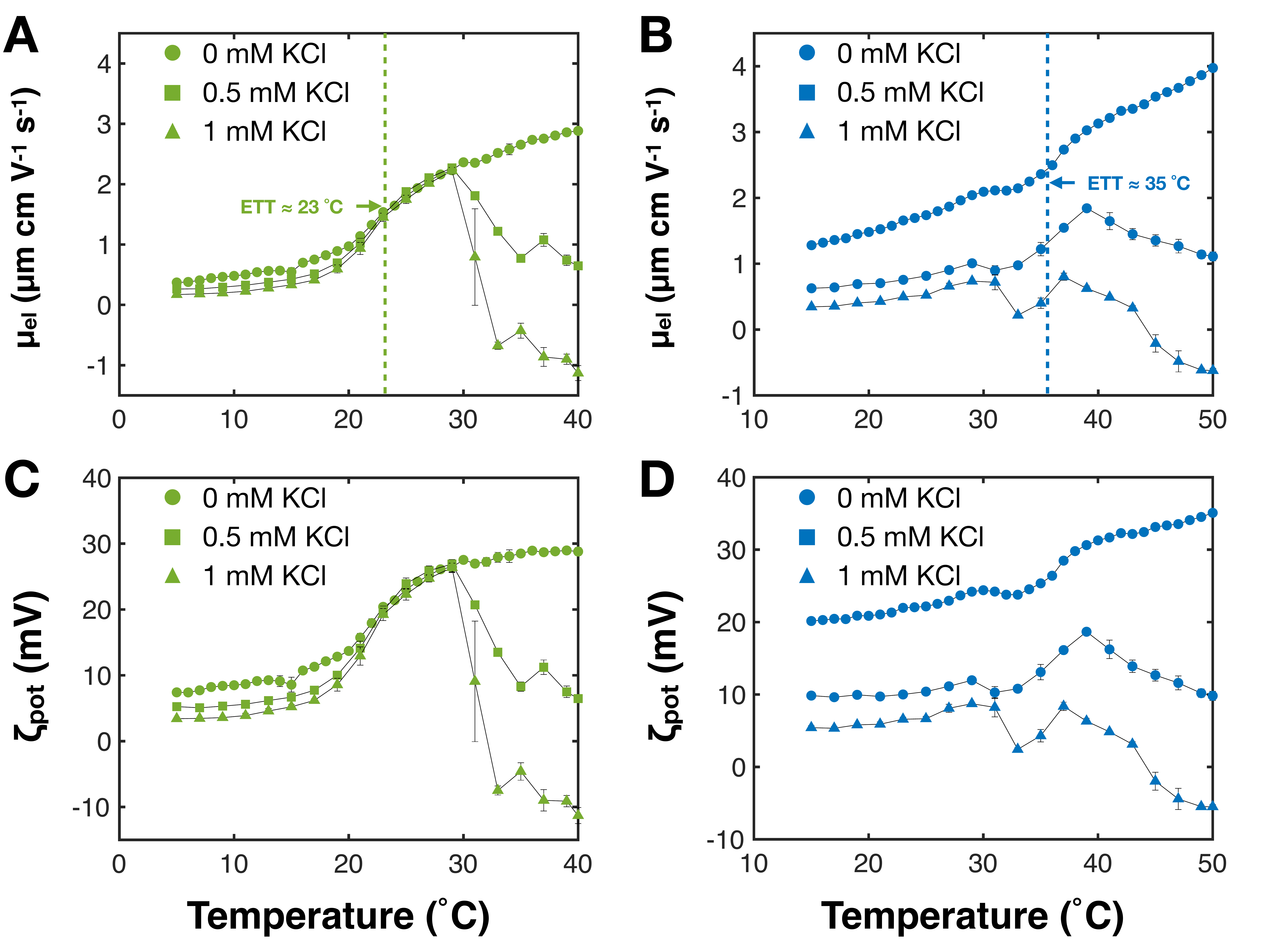}
\caption{(A, B) Electrophoretic mobility and (C, D) zeta potential as a function of temperature for (A, C)  pNIPAM-co-DEAAM and (B, D) pNIPAM microgels at different KCl concentrations. The dashed lines show the ETT of the microgels.
}
\label{fig:figure_MobZeta}
\end{figure}

\begin{figure}[ht!]
\includegraphics[width=.90\textwidth ]{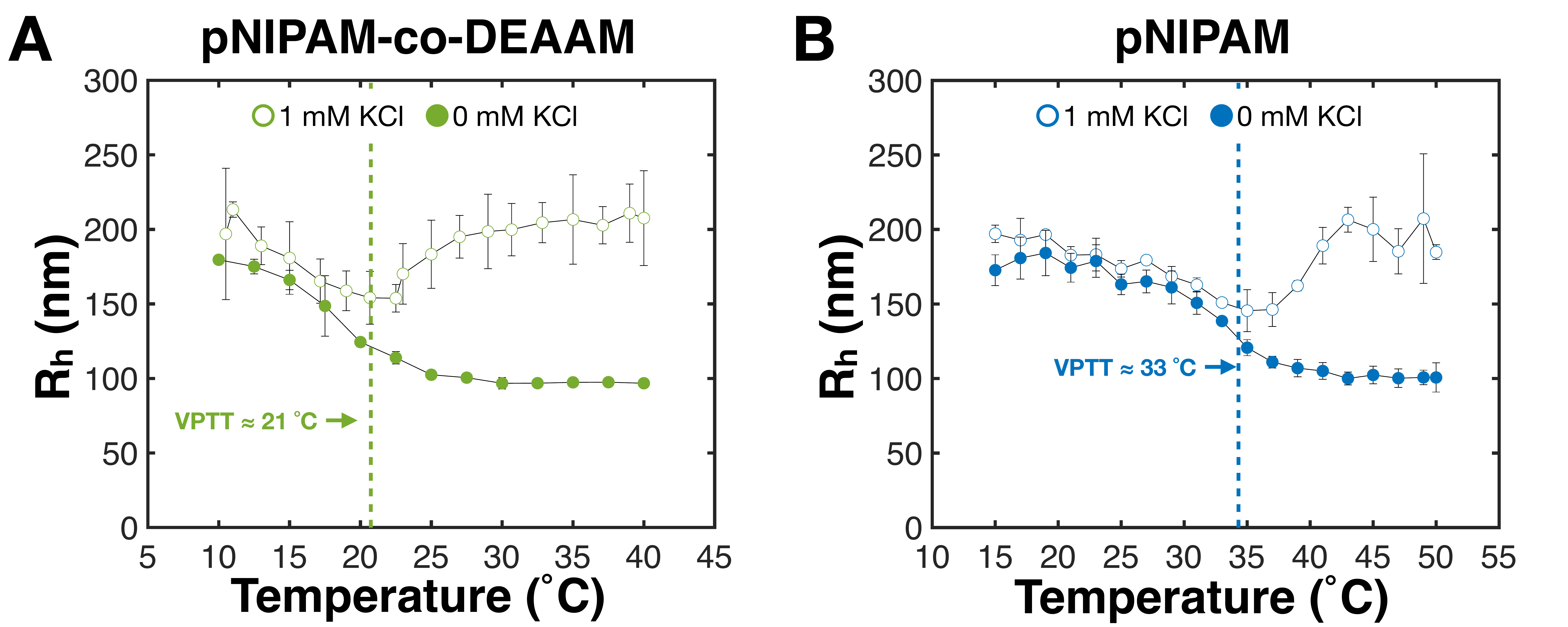}
\caption{Temperature-dependent swelling of the microgels measured with dynamic light scattering in 0 and 1~mM KCl solution. Measurements were conducted with a NanoZS Zetasizer at a scattering angle of 173$^\circ$. (A) pNIPAM-co-DEAAM microgels and (B) pNIPAM. 
}
\label{fig:figure_DLS_1mM_KCl}
\end{figure}

\newpage
\section{Force-Distance curves}

\begin{figure}[ht!]
\includegraphics[width=0.7\textwidth]{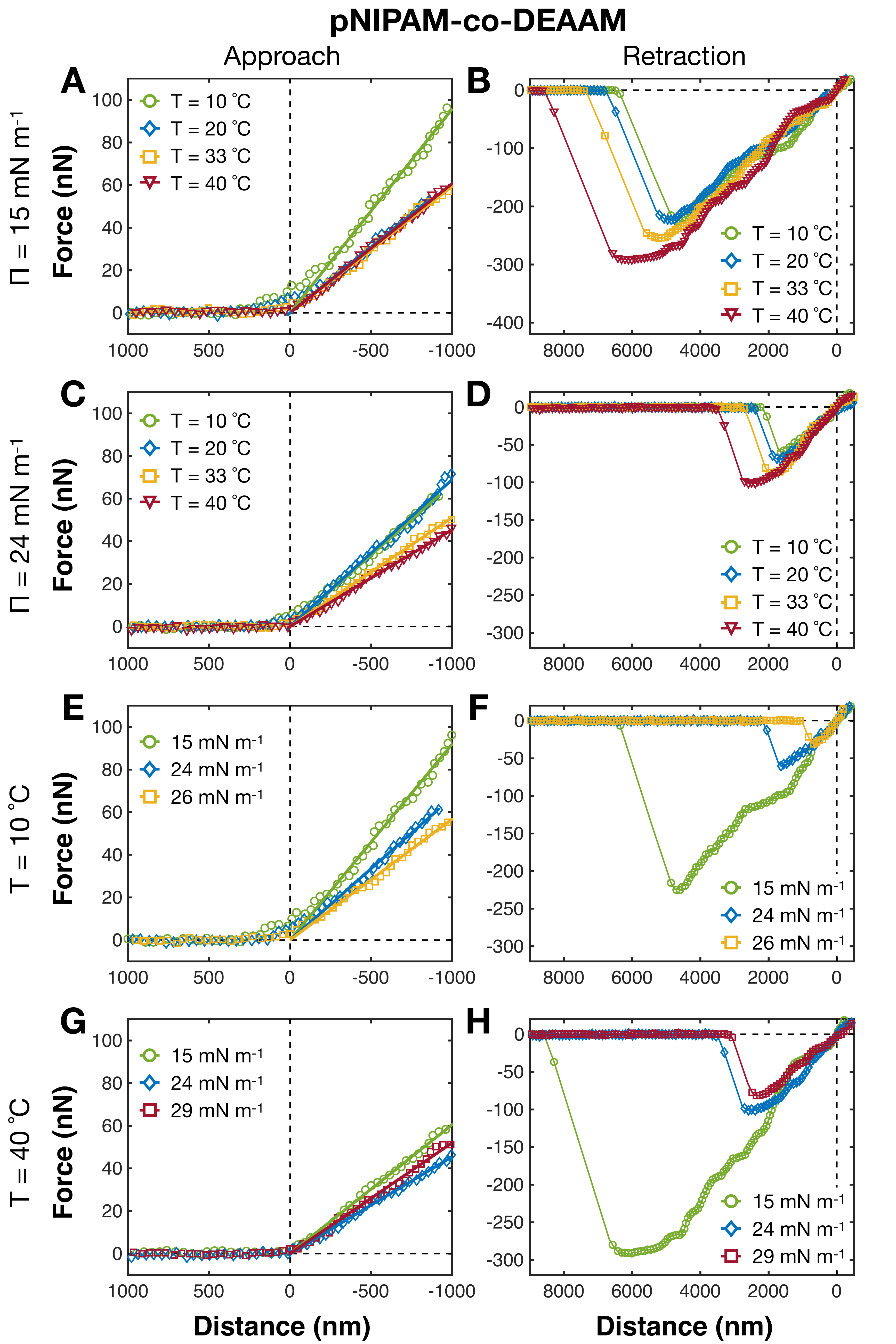}
\caption{Examples of force-distance curves of pNIPAM-co-DEAAM microgel monolayers at the air-water interface measured with hydrophilpos probe from the aqueous phase. The probe has diameter of 6.84 $\mu$m. Approach (left) and retraction (right) force-distance curves. (A,B) F-D curves as a function of temperature. (C,D) F-D curves as a function of compression for T = 20~$^\circ$\,C. (E,F) F-D curves as a function of compression for T = 40 $^\circ$\,C.}
\label{fig:figure_FD_SB103c}
\end{figure}

\begin{figure}[ht!]
\includegraphics[width=.90\textwidth]{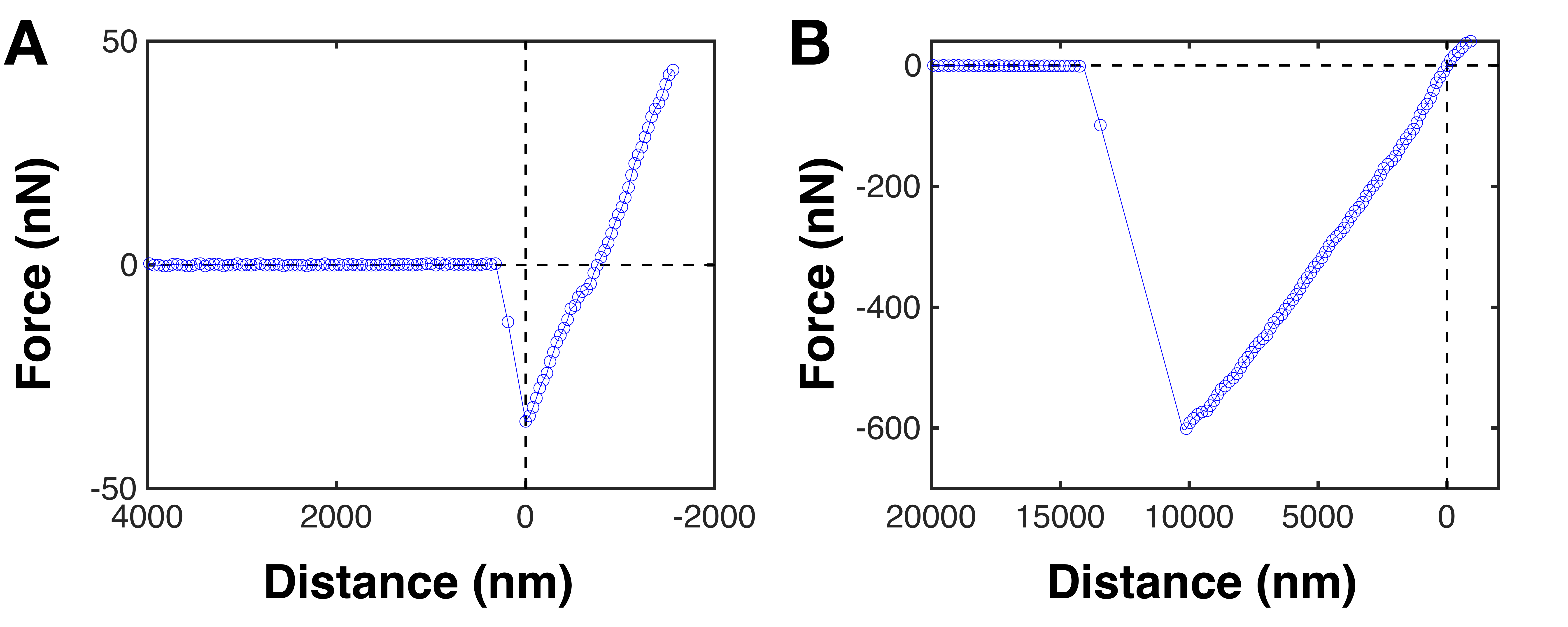}
\caption{Force distance curves of the bare air-water interface measured with a hydrophilized silica particle as probe from the aqueous phase. (A) Approach force curves, and (B) retract force curves at T~=~20~$^\circ$C. The black vertical dashed lines represent point where the prob is fully is in contact with the monolayer and the slope linear. Black horizontal dashed lines show zero force.
}
\label{fig:figure_FD_bare_Water}
\end{figure}